**Effects of Transit Signal Priority on Traffic Safety: Interrupted Time Series Analysis of Portland, Oregon, Implementations**

**Yu Song***
Department of Civil and Environmental Engineering
Traffic Operations and Safety Laboratory
University of Wisconsin-Madison
1415 Engineering Dr.
Madison, WI 53706
E-mail: yu.song@wisc.edu
*Corresponding Author

**David Noyce**
Department of Civil and Environmental Engineering
Traffic Operations and Safety Laboratory
University of Wisconsin-Madison
1415 Engineering Dr.
Madison, WI 53706

**ABSTRACT**

Transit signal priority (TSP) has been implemented to transit systems in many cities of the United States. In evaluating TSP systems, more attention has been given to its operational effects than to its safety effects. Existing studies assessing TSP's safety effects reported mixed results, indicating that the safety effects of TSP vary in different contexts. In this study, TSP implementations in Portland, Oregon, were assessed using interrupted time series analysis (ITSA) on month-to-month changes in number of crashes from January 1995 to December 2010. Single-group and controlled ITSA were conducted for all crashes, property-damage-only crashes, fatal and injury crashes, pedestrian-involved crashes, and bike-involved crashes. Evaluation of the post-intervention period (2003-2010) showed a reduction in all crashes on street sections with TSP (-4.5%), comparing with the counterfactual estimations based on the control group data. The reduction in property-damage-only crashes (-10.0%) contributed the most to the overall reduction. Fatal and injury crashes leveled out after TSP implementation but did not change significantly comparing with the control group. Pedestrian and bike-involved crashes were found to increase in the post-intervention period with TSP, comparing with the control group. Potential reasons to these TSP effects on traffic safety were discussed.

Key words: transit signal priority; interrupted time series analysis; traffic safety; crash.

**INTRODUCTION**

Transit signal priority (TSP) has been implemented to transit systems in many cities of the United States. TSP is a type of transit preferential treatment installed at signalized intersections. The purpose of TSP is to improve transit operational performance through adjusting signal timing and providing approaching transit vehicles longer or earlier green lights (*1,2*). For bus TSP implementations in cities of United States, the typical travel time saving ranges from 8% to 12% (*3*). TSP does increase cross-street traffic delays to some extent but can be kept to a minimum by using GPS-based or adaptive TSP systems (*4*). Despite the well-understood significant improvement of operational performance brought by TSP, its safety effects are usually overlooked by transit agencies in the project development process. As the working process of TSP involves signal timing adjustments, such a treatment can directly or indirectly affect traffic safety performance of signalized intersections and their adjacent roadway segments (*5–9*).

Existing studies reported mixed results regarding TSP's effects on traffic safety. Some implementations were found to be correlated with increase in number of crashes, and some others were found to correlate with reduction in number of crashes. Existing studies have been focusing on (general or transit-involved) vehicle crashes with different severities but have not given enough attention to pedestrian and bike crashes. If data availability allows, pedestrians and bikes should be considered in road safety assessments. Pedestrians and bicyclists are the most vulnerable users in transportation systems, with less protection, lower weights and lower speeds, comparing with motor vehicles (*10*). Also, most of pedestrians' and bicyclists' activities take place in urban areas where motor vehicle activities and transit services also take place in (*11*). In this study, these two groups of road users were considered.

In terms of study methods, cross-sectional and longitudinal study methods have been used in existing studies of TSP safety effects. Interrupted time series analysis (ITSA) is a longitudinal study method. It is widely used in medical, public health, and public policy research, and has been proved by a great number of existing studies to be powerful in not only providing estimation of effectiveness but also revealing the detailed changing process (including the immediate change in level, a rate of changing over time, and the amount of change) before and after interventions (*40*). In traffic safety field, time series analytical methods like ITSA have not been frequently applied due to insufficient knowledge on selecting and applying appropriate modeling techniques, and difficulties in the gathering of data with appropriate structures. However, with the advantages of this method, and the improvement in the availability of longitudinal safety data, the application of time series analysis methods in traffic safety research and practices is promising (*12*). For this study, the available crash data set has a longitudinal structure suitable for an ITSA, thus ITSA was selected as the assessment method.

The data used in this study were obtained from Portland, Oregon, which is one of the very first cities that implemented TSP in the United States. Thirteen TSP corridors in Portland, which had TSP deployed in June to October 2002, were selected as a "treated group" for the assessment. Street sections in Portland without TSP were matched to the treated sections, forming a "control group" for a controlled assessment. Traffic crash data from 1995 to 2010 was aggregated by month for the treated and control groups. Models were developed for all crashes, property-damage-only (PDO) crashes, fatal and injury (FI) crashes, pedestrian-involved crashes, and bike-involved crashes, respectively.

## LITERATURE REVIEW

The safety aspect of TSP's effects did not receive a lot of attention until recent years. There are now several academic publications on this topic. However, these prior studies only covered TSP implementations in three cities of three countries. TSP, as a system, is usually designed and configured based on transit agencies operational demand, traffic signals' capabilities, and local traffic conditions. Thus, different TSP implementations are considered to perform somewhat differently in operational and safety perspectives. As reflected by the existing studies, TSP affected the traffic safety in three metropolitan areas differently.

Two evaluations of TSP implementations in Toronto, Canada, reported correlations between TSP and increase in traffic crashes. Shahla et al. studies Toronto's bus and streetcar TSP system using a cross-sectional study method with negative binomial regression models (*14*). Four years (1999–2003) of crash data were analyzed. Modeling results showed that the implementation of TSP was correlated with the increase of crashes at signalized intersections. In terms of all crashes, the implementation of streetcar TSP brought a 31% increase in crashes, while bus TSP brought a 28% crash increase. From two different models, TSP was found related to a 25% or 51% increase in the combined streetcar/bus-involved crashes. Li et al. studied the same TSP system using Vissim micro-simulation and Surrogate Safety Assessment Model (SSAM) (*15*). Five years (2006-2010) of data were used in the analysis. Intersection-level vehicular conflicts were reported from the simulation models and were converted to crashes using a negative binomial regression models reflecting the relationship between the numbers of different types of crashes and the numbers of different types of conflicts. The analysis results showed that removing TSP from the existing, TSP-enabled, intersections could lead to a reduction in all crashes by as much as 1.6%, a 2.9% reduction in angle crashes, a 1.9% reduction in rear-end crashes, and a 2.1% reduction in side-swipe crashes. However, with the removal of TSP, changes in transit-involved crashes could range from a reduction of 0.1% to an increase of 1.6%. These results indicated that the enabling of TSP would increase traffic crash numbers.

Three studies of TSP implementations in Melbourne, Australia, showed that TSP helped reduce number of crashes. Goh et al. explored the traffic safety impacts of bus priority treatments including bus lanes and TSP using Empirical Bayes (EB) before-after analysis method (*16*). Particularly, in terms of TSP, the results showed an 11.1% reduction in expected FI crashes in the "after" period with TSP implemented, compared with the "after" period crash frequency estimates without TSP. Another cross-sectional evaluation of TSP's effects on bus-involved crashes by Goh et al., using negative binomial regression and neural network modeling methods, showed that Melbourne TSP, in collaboration with non-TSP transit preferential treatments (e.g., bus lane, stop relocation), helped reduce 53.5% of bus-involved crashes (*17*). Naznin et al. assessed streetcar TSP at 29 intersections using EB before-after study method (*18*). A statistically significant reduction of 13.9% in streetcar-involved crashes was found after the implementation of TSP. A disaggregate-level simple before–after analysis also indicated reductions in total and FI crashes as well as vehicle, pedestrian, and motorcycle-involved crashes.

One study of TSP implementations in the Seattle metropolitan area, United States, reported reduction in number of traffic crashes after TSP deployment. Song and Noyce assessed the effects of TSP implemented with King County Metro Transit's RapidRide enhanced bus system in 2010-2014, using EB before-after study (*19*). The study results showed that the implementation of TSP was significantly correlated with a 13% reduction in all crashes, a 16% reduction in property-damage-only crashes, and a 5% reduction in fatal and injury crashes.

From reviewing prior studies on TSP's safety effects, the following three potential improvements were identified for this and future studies to make on this topic.

- Prior studies investigated TSP implementations in only three metropolitan areas, evaluations should be carried out to get a better idea of TSP's safety performance in different geographical areas and under different traffic/transit operational contexts.
- Prior studies have not been able to pay enough attention to road users like pedestrians and bicyclists, which are the most vulnerable groups that can potentially be influences by changes to traffic control devices.
- Prior studies yielded mixed results about TSP's effects on safety, thus more studies are needed to better understand what made TSP affect traffic safety differently.

**DATA**
**Description**
Data used for this study were from Portland, Oregon, a city with a long history of using TSP in its transit systems. Street sections with TSP systems activated from June to December 2002 were selected as the study sites (i.e., treated group for the ITSA). Information of the treated group is listed in Table 1. Locations and limits of the treated group are mapped in Figure 1. For a controlled ITSA, a control group consisting of street sections, similar with the treated group in terms of geometric and traffic features, but without TSP, was also needed. TSP implementation information and street feature data for both the treated and control groups were provided by Portland Bureau of Transportation (PBOT). Street sections for the control group were selected using a propensity score matching (PSM) method which will be explained later in the text.

Portland crash data from 1995 to 2010 were obtained from Oregon Traffic Safety Data Archive (OrTSDA) maintained by Portland State University and Oregon Department of Transportation (ODOT). Information of crashes related to the selected street sections were extracted using ArcGIS geoprocessing tools, with a buffer distance of 100 feet to capture crashes within the effective area of TSP. Crashes were then aggregated to each street section based on its primary street ID in the attribute table.

Portland traffic data from 1995 to 2010 were obtained from PBOT. Traffic counts were extracted using ArcGIS geoprocessing tools, with a buffer of 5 feet considering coordinate errors. The aggregated traffic data set had multiple missing values, thus could not be used in the ITSA as an exposure variable, so number of lanes was used as an exposure in the ITSA instead. Despite the missing values in traffic volume data, averaged traffic volumes over the pre-intervention months could be calculated and were sufficient to be used in the PSM for control group selection.

**Table 1  Detailed Information of Treated Group and Control Group Street Sections**

**Treated Group**

| # | Name | Limit | Lanes | One-way[a] | Median[b] | Length[c] | AADT[d] | TSP-On Date |
|---|------|-------|-------|---------|--------|--------|-------|-------------|
| 1 | NE Sandy Blvd | NE 96th Ave – NE 47th Ave | 4 | 0 | 0 | 2.75 | 19967 | 6/2002 |
| 2 | NE Sandy Blvd | NE 39th Ave – NE 16th Ave | 4 | 0 | 0 | 1.40 | 22411 | 6/2002 |
| 3 | E Burnside St | NE 9th Ave – MLK Jr. Blvd | 3 | 1 | 0 | 0.24 | 28866 | 6&7/2002 |
| 4 | NE & SE 82nd Ave | SE Flavel St – NE Killingsworth St | 4 | 0 | 0 | 6.36 | 28302 | 10–12/2002 |
| 5 | NE Alberta St | NE 15th Ave – NE MLK Jr. Blvd | 2 | 0 | 0 | 0.51 | 9613 | 11/2002 |
| 6 | SE Powell Blvd | SE 162nd Ave – SE 104th Ave | 2 | 0 | 0 | 2.93 | 22009 | 12/2002 |
| 7 | SE Powell Blvd | SE 92nd Ave – SE 52nd Ave | 4 | 0 | 50 | 1.96 | 29359 | 8&12/2002 |
| 8 | SE Powell Blvd | SE 50th Ave – SE Milwaukie Ave | 4 | 0 | 0 | 2.13 | 43621 | 10/2002 |
| 9 | NE Broadway St | NE 24th Ave – NE 7th Ave | 3 | 1 | 0 | 0.78 | 17192 | 6&7/2002 |
| 10 | NE Weidler St | NE 7th Ave – NE 21st Ave | 3 | 1 | 0 | 0.61 | 16753 | 7/2002 |
| 11 | SE Hawthorne Blvd | SE 27th Ave – SE 40th Ave | 4 | 0 | 0 | 0.81 | 17969 | 8/2002 |
| 12 | SE Hawthorne Blvd | SE 41st Ave – SE 50th Ave | 2 | 0 | 0 | 0.42 | 13384 | 8/2002 |
| 13 | SE Foster Rd | SE 52nd Ave – SE 82nd Ave | 4 | 0 | 0 | 1.72 | 21782 | 10/2002 |

**Control Group**

| # | Name | Limit | Lanes | One-way | Median | Length | AADT | Times Matched |
|---|------|-------|-------|---------|--------|--------|------|---------------|
| 1 | N Albina Ave | NE Lombard – N Prescott | 2 | 0 | 0 | 1.51 | 5255 | 12 |
| 2 | NE Columbia Blvd | NE 33rd Ave – NE 47th Ave | 4 | 0 | 0 | 0.53 | 28158 | 1 |
| 3 | NE Lombard St | N Williams Ave – NE MLK Jr. Blvd | 2 | 0 | 1 | 0.26 | 21771 | 1 |
| 4 | SE 17th Ave | SE Tenino St – SE Ochoco St | 2 | 0 | 0 | 0.35 | 16299 | 12 |
| 5 | SE Division St | SE 11th Ave – SE 17th Ave | 2 | 0 | 0 | 0.30 | 12503 | 12 |
| 6 | SE Division St | SE 61st Ave – SE 70th Ave | 2 | 0 | 0 | 0.42 | 18025 | 1 |
| 7 | SE Division St | SE 96th Ave – SE 111th Ave | 4 | 0 | 0 | 0.68 | 43174 | 12 |
| 8 | SE Division St | SE 137th Ave – SE 142nd Ave | 4 | 0 | 0 | 0.27 | 41805 | 12 |
| 9 | SE Division St | SE 148th Ave – SE 162nd Ave | 4 | 0 | 0 | 0.49 | 33660 | 1 |
| 10 | SE Foster Rd | SE 128th Ave – SE 136th Ave | 3 | 0 | 0 | 0.30 | 21171 | 1 |

Note: a. 0 = Two-way, 1 = One-way;

b. Percentage of street section with median (%);

c. Street section length (mi);

d. Average value (across pre-intervention period 1995-2001) of Annual Average Daily Traffic (AADT) in vehicles per day (vpd);

E: East, N: North, NE: Northeast, SE: Southeast; Ave: Avenue, Blvd: Boulevard, Hwy: Highway, Rd: Road, St: Street;

MLK Jr.: Martin Luther King Junior.

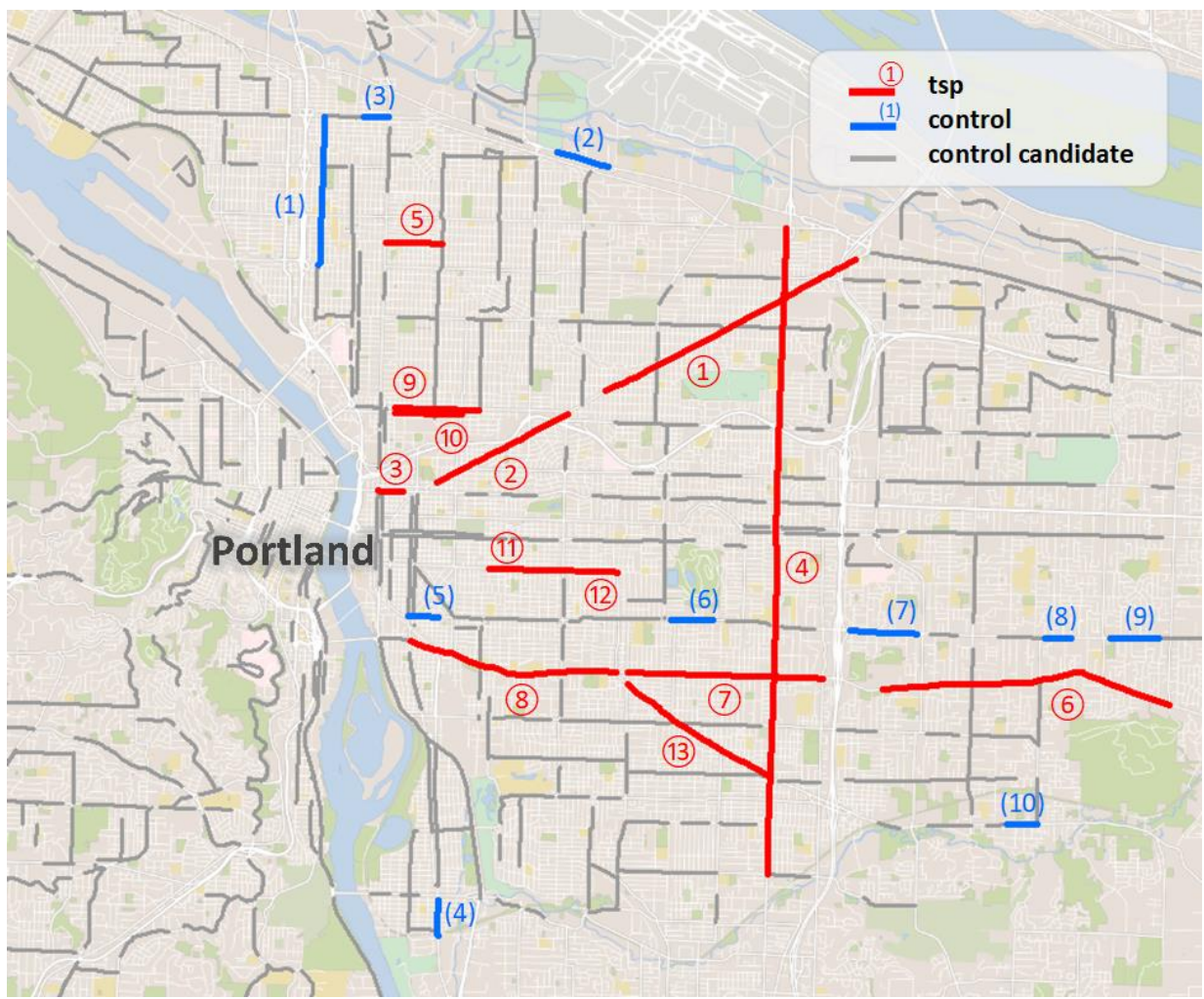

**Figure 1  Treated Group and Control Group Street Sections**

Bus routes and schedule data were provided by TriMet, the public transit agency for the Portland metropolitan area. Signalized intersection data were obtained from the Portland Maps Open Data website (https://gis-pdx.opendata.arcgis.com/). These data were also used in the PSM process.

**Selection of Control Group using Propensity Score Matching**

PSM is a statistical matching method, developed by Rosenbaum and Rubin, for selecting control groups in observational studies (*20*). The method involves a regression model estimating the conditional probability of an entity (e.g., street section, intersection) being selected to apply a treatment, given a vector of observed covariates. For an observational study like this one, the treated group and control group are not picked randomly as in an experiment, thus to avoid possible selection bias in choosing the control group and the bias in estimating treatment effects, a statistical matching process should be performed (*21*). There are three assumptions for PSM (*22*):

- Stable unit treatment value assumption (SUTVA), stating that there is no interference between entities. For this study, crashes were aggregated to each selected street section based on their primary street IDs in the attribute table, thus each crash was assigned to only one street section. There was no overlap in crashes between any two street sections.
- Positivity, stating that every entity has a non-zero probability of being assigned to receive or to not receive the treatment. In terms of this study, all street sections were estimated to have a probability of receiving the TSP treatment in the range of (0,1), thus there was positive chances for a street section to receive or not receive the treatment.

- Unconfoundedness, stating that the treatment assignment mechanism is unconfounded if the treatment assignment is conditionally independent of the potential outcomes. For this study, all street sections included in the PSM have bus routes operating on them, and all available street geometric features, traffic features, and crash history (during pre-intervention period) were used as covariates in the PSM.

Binary logit regression has been used in previous road safety studies (*21–25*). Thus, for this study, a binary logit regression model was developed to estimate the propensity scores for the treated entities and candidate control group entities:

$$
\begin{aligned}
\Pr(Y_i = 1) = \{1 \\
+ \exp[-34.30 + 3.18\ln(AADT_i) + 0.32(Oneway_i) - 0.77(Lanes_i) \\
+ 4.44(Median_i) - 2.08(BusRoutes_i) + 0.57(MajorFrequency_i) \\
+ 0.10(SecondaryFrequency_i) + 0.28(SignalDensity_i) + \ln(Length_i)]\}^{-1} \quad (1)
\end{aligned}
$$

Where:

$\Pr(Y_i = 1)$ = propensity score reflecting the probability of a street section being selected for TSP implementation;

$AADT_i$ = annual average daily traffic value averaged across years in pre-intervention period (1995-2001) of street section i (vpd);

$Oneway_i$ = indicator of one-way street section, 0 = not one-way, 1 = one-way;

$Lanes_i$ = number of lanes on street section i;

$Median_i$ = median ratio (median length / section length) of street section i;

$BusRoutes_i$ = number of bus routes (two directions) on street section i;

$MajorFrequency_i$ = Combined two-direction frequency (buses/h) of Crosstown, Reginal Trunk, and Urban Radial bus routes that operate along street section i;

$SecondaryFrequency_i$ = Combined two-direction frequency (buses/h) of Employer Feeder, Feeder, Peak Express, and Secondary Feeder bus routes that operate along street section i;

$SignalDensity_i$ = density of signalized intersections of street section i (signals/mi);

$Length_i$ = length of street section i (mi), as an exposure variable.

There are several different matching methods for PSM such as nearest-neighbor matching, K-nearest neighbor matching, radius matching, kernel matching, and Mahalanobis matching (*26*). In this study, 5-nearest neighbor matching with replacement was used. To each treated street section, five control street sections having the smallest difference in propensity score, comparing with the treated section, were identified as matches. As replacement was used, the 65-section control group was formed by 10 distinct street sections, as listed in Table 1 and mapped in Figure 1.

The final data set for ITSA had the crashes of the treated group and the control group organized by month, from January 1995 to December 2010 (16 years, or 192 months, in total). In addition to the data of all crashes, data of PDO, FI, pedestrian-involved, and bike-involved crashes were also extracted from the raw data sets and organized by month for separate analyses. Over the 192-month study period, the treated group had a total of 22065 crashes (55% PDO, 45% FI, 3% pedestrian, and 2% bike), and the control group had a total of 18795 crashes (54% PDO, 46% FI, 2% pedestrian, and 2% bike). Apart from crash numbers, other variables were also added to the data set for various ITSA modeling needs, which are explained in the Methodology section.

To ensure that the selected control group and the treated group are comparable, the sample odds ratios between the two groups annual crash numbers during the pre-intervention period, were calculated (*27*). As shown in Table 2, the odds ratios for all crashes, PDO, and FI crashes were very close to 1, which indicates good comparability between the two groups. The pedestrian and bike-involved crashes had much fewer data than the other categories, thus had higher odds ratios and wider 95% confidence intervals.

**Table 2  Pre-Intervention Sample Odds Ratio Between Treated and Control Group**

| Crash Type | Mean Odds Ratio | Standard Error | 95% Confidence Interval | |
|:---:|:---:|:---:|:---:|:---:|
| **All** | 1.03 | 0.03 | 0.97 | 1.08 |
| **PDO** | 1.03 | 0.04 | 0.95 | 1.11 |
| **FI** | 1.02 | 0.08 | 0.86 | 1.18 |
| **Pedestrian** | 1.76 | 0.72 | 0.36 | 3.17 |
| **Bike** | 2.30 | 1.48 | -0.59 | 5.20 |

## METHODOLOGY

ITSA is a method widely used in medical and public health studies, as well as public policy research (28–33). It has been proven by a great number of existing research practices to be a powerful tool for longitudinal intervention assessments, which is able to not only provide estimations of effectiveness but also reveal the detailed changes of the measures of effectiveness (MOE) before and after the interventions (30,33). Time series analytical methods have been frequently applied in traffic operational studies for flow modeling and prediction (34–36). However, in traffic safety area, these methods have yet been widely applied, and ITSA applications are especially rare (12). Despite this fact, there are several existing traffic safety studies evaluating the effectiveness of a (infrastructural or political) treatment using ITSA (37–40).

An interrupted time series design has observations collected repeatedly with equal time intervals, from a panel of entities, before and after a treatment intervention. The data trajectory has the possibility of being interrupted by the intervention and thus change. An interrupted time series model compares the level of the post-intervention predictions had no intervention is introduced against the post-intervention estimations that occur after the intervention is introduced. ITSA has some strengths in terms of estimating the effects of a treatment:

- Comparing with simple before-after (or pre-post) design, ITSA controls for the effect of secular trends in a time series of the outcome MOE (30).
- In addition to an estimation of the intervention's before-after effectiveness, ITSA gives more detailed illustrations of changes over time (40).
- ITSA provides clear and easy-to-interpret graphical results (30).

When data of a matched control group is available, a controlled ITSA can be carried out and provide enhanced analysis addressing issues of confounding omitted factors (43). In this study, both single-group and controlled ITSA were carried out to assess the effects of TSP on all crashes, PDO crashes, FI crashes, pedestrian-involved crashes, and bike-involved crashes.

Modeling methods for ITSA include autoregressive integrated moving-average (ARIMA) intervention and time series regression intervention models with autoregressive. Autoregression is a stochastic process used in statistical calculations where a value from a time series is regressed on previous values from that same time series, to accommodate the impact (autocorrelation) between temporally adjacent observations. In this study, time series regression intervention models with autoregressive errors were used for the ITSA to address the possible autocorrelation between one estimation's error term and its previous error terms. In addition to the autoregressive error term, the ITSA models used in this study also considered possible seasonality existing in the time series by adding month indicators. Seasonality is a regular and predictable pattern in a time series that appear once every period, which, in the case of the crash data used here, can be a month, two months, or several months. Regression to the mean (RTM) is another potential issue for single-group ITSA, but the issue is not significant in the case of this study. For pre-post intervention study designs like this one, RTM is particularly a concern when the study targets "high risk" groups (e.g., safety treatments applied to high-crash sites). This study is a pre-post intervention study, but the effectiveness measure, number of crashes, is not a target measure of the treatment, TSP, when it was implemented. The best solution to RTM is to carry out a controlled analysis using an appropriately selected control group. To address time-varying confounders that are either unmeasured or unknown, and to overcome RTM effects, a controlled ITSA was carried out in addition to the single-group ITSA, data from the PSM matched control group were added to the analysis. All analyses in this study were carried out in R.

For the single-group ITSA, regression models of the following form were developed:

$$Y_t = \beta_0 + \beta_1 time_t + \beta_2 level_t + \beta_3 trend_t + \alpha_1 M_1 + \alpha_2 M_2 + \cdots + \alpha_n M_n + \epsilon_t \qquad (2)$$

Where:

$Y_t$ = aggregated outcome variable (number of crashes) of the treated group, measured at each equally spaced time point t (month);

$time_t$ = the time (month) since the start of the study (January 1995);

$level_t$ = TSP intervention rate (0 to 1);

$trend_t$ = the time (month) after the intervention, the interaction between $time_t$ and $level_t$;

$M_1 \ldots M_n$ = month indicators (n = 1 to 11), reflecting seasonality of the crash outcome, only the months with significant coefficients were kept for each model;

$\epsilon_t$ = error term.

The coefficients to be assessed to determine TSP's effects are:

$\beta_0$ = intercept, reflecting the initial level of the outcome variable;

$\beta_1$ = slope of the outcome variable in pre-intervention phase;

$\beta_2$ = immediate change following the introduction of intervention, the immediate treatment effect;

$\beta_3$ = difference between pre-intervention and post-intervention slopes of the outcome, the treatment effect over time.

Note that the outcome measure used in this study is the number of crashes per 100 lane-miles, considering the length and number of lanes of the treated group as exposures. For the level variable, different from the binary level variable representing whether or not an intervention was implemented, here the variable is treated as a continuous, rate of intervention, as the implementation of TSP was carried out progressively from June to December 2002. The rate was calculated by dividing the number of intersections in the treated group with TSP installed and activated to the total number of intersections in the treated group. This type of variable treatment has been applied in before in a traffic safety study by Huitema et al. (*40*).

Autoregressions was incorporated in the error term:

$$\epsilon_t = \phi_1 \epsilon_{t-1} + \phi_2 \epsilon_{t-2} + \cdots + \phi_p \epsilon_{t-p} + \omega_t \qquad (3)$$

Where:

$\epsilon_t$ = error term of current observation;

$\phi_i$ = correlation coefficient between the current error term and its i[th] previous error term;

$\epsilon_{t-i}$ = error term of the i[th] previous observation;

$\omega_t$ = error term following usual assumptions about regression errors, independent and identically distributed *N(0, $\sigma^2$)*.

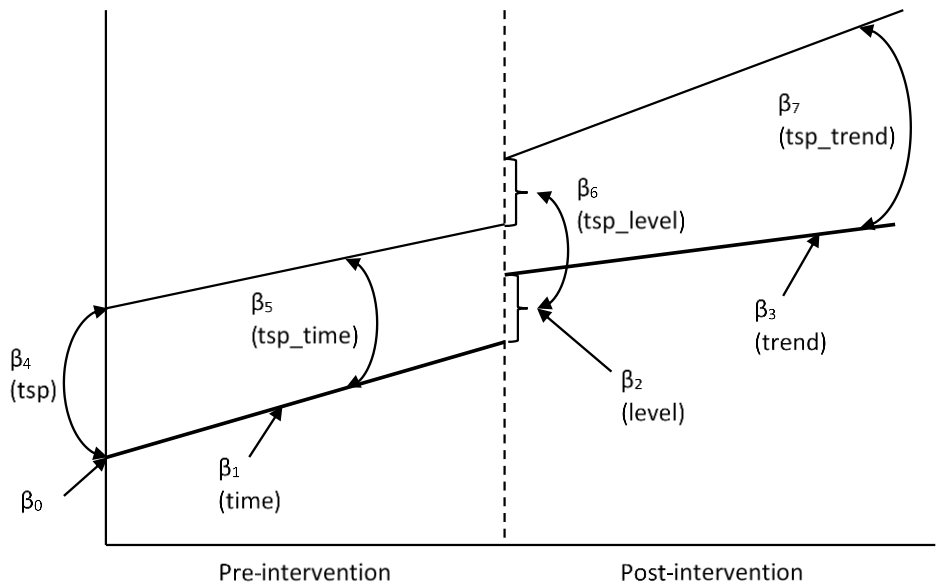

**Figure 2  Illustration of Coefficients of a Controlled ITSA Model (*43*)**

For each model, the order of autoregression (i.e., autoregressive lag), p, was determined using Durbin-Watson tests in collaboration with autocorrelation function (ACF) and partial autocorrelation function (PACF) plots (*41,42*). After this model checking and adjusting process, a sensitivity analysis was then performed for each model using analysis of variance (ANOVA) to test the impacts of varying the autoregression and the moving average lags. Quantile-quantile (q-q) plots were used to check the normality of the models' residuals. All the final models presented in this paper were tested in such a way and ensured to be relatively optimal.

For the controlled ITSA, the following form of regression models were used:

$$Y_t = \beta_0 + \beta_1 time_t + \beta_2 level_t + \beta_3 trend_t + \beta_4 tsp_t + \beta_5 tsp\_time_t + \beta_6 tsp\_level_t + \beta_7 tsp\_trend_t + \alpha_1 M_1 + \alpha_2 M_2 + \cdots + \alpha_n M_n + \epsilon_t \qquad (4)$$

Where:

$Y_t$ = aggregated outcome variable (number of crashes) of the treated and control groups, measured at each equally spaced time point t (month);

$time_t$, $level_t$, and $trend_t$ are as defined in Equation (2);

$tsp_t$ = indicator to identify groups, 1 = treated group, 0 = control group;

$tsp\_time_t$, $tsp\_level_t$, and $tsp\_trend_t$ = interactions between $tsp_t$ and $time_t$, $level_t$, and $trend_t$.

The coefficients to be assessed are as follows and are visually illustrated in Figure 2:

$\beta_0$, $\beta_1$, $\beta_2$, and $\beta_3$ = coefficients describing the initial level, slope, level change, and trend change of the control group crashes;

$\beta_4$ = difference in the start level of the crashes between the treated group and the control group;

$\beta_5$ = difference in the slope of crash number change in the pre-intervention phase between the treated group and the control group;

$\beta_6$ = difference in the immediate effect of intervention between the treated group and the control group;

$\beta_7$ = difference in the pre and post-intervention slope differences between the treated group and the control group.

## RESULTS AND DISCUSSION

Single-group and controlled ITSA were conducted for all crashes, PDO crashes, FI crashes, pedestrian-involved crashes, and bike-involved crashes. The 90th month, June 2002, was the initial month of the TSP intervention. This section presents the ITSA results in forms of table and chart, and provides the interpretations of these results, to get an understanding of how the TSP implementations in Portland, Oregon affected traffic safety.

To make the outcomes comparable, and the dependent variables have more readable and easier to interpret coefficients, the outcome measures were standardized and rescaled to the number of crashes per 100 lane-miles (i.e., 100 * the number of crashes per mile per lane). The comparisons and interpretations in this section are based on this scale of the outcome measure.

### Single-Group Analysis

The annual crash numbers of TSP treated street sections followed a reducing trend during the 16-year period from 1995 to 2010. The number of all crashes reduced by 21% (1533 to 1216), where PDO crashes reduced by 21% (745 to 587) and FI crashes reduced by 20% (788 to 628). However, pedestrian-involved crashes increased by 13% (39 to 44), and bike-involved crashes increased by 13% (24 to 27).

Visual displays of the single-group ITSA regressions are presented in Figure 3. The solid lines represent estimated trend from actual crash data points, and the dashed lines represent the predicted counterfactual trend, assuming crash numbers would follow the pre-intervention trend had TSP not been implemented. To present a simpler visualization for easier comparison of the estimations, the seasonality effects were excluded in the visual displays. Detailed model coefficients and parameters are listed in Table 3. In terms of TSP intervention's effects, the coefficient estimations for "level" and "trend" reflect the immediate (difference in crash number) and temporal (difference in slope) effects on monthly crashes. Overall, comparing with the pre-intervention trend, there is a slight reduction in both the level (-0.44, 95% CI [-21.92, 21.04]) and trend (-0.11, 95% CI [-0.48, 0.26]) of all crashes after the intervention of TSP, but not significant. The PDO crashes had an insignificant drop in level (-2.77, 95% CI [-17.85, 12.32]) but a significant reduction in the post-intervention trend (-0.52, 95% CI [-0.78, -0.26]). The FI crashes had an insignificant increase in level (2.42, 95% CI [-7.86, 12.70]) and a significant increase in trend (0.40, 95% CI [0.22, 0.57]) in the post-intervention period, comparing with the pre-intervention period. The pedestrian-involved crashes had an insignificant reduction in level (-1.20, 95% CI [-2.84, 0.43]) and a significant increase in trend (0.04, 95% CI [0.01, 0.07]). The bike-involved crashes had an insignificant increase in level (0.10, 95% CI [-1.03, 1.23]), and no change in trend (0.00, 95% CI [-0.02, 0.02]).

As mentioned in previous sections of this paper, apart from the effects of the intervention of interest, there could be time-varying confounders that are either unmeasured or unknown. Therefore, other than giving a rough idea of how the MOE changed after the intervention, a single-group ITSA cannot provide an accurate estimation of TSP's safety effects. Conducting a controlled ITSA is necessary to reduce uncertainties from the estimations and overcome the potential RTM effects in single-group pre-post analysis. The findings of this study are mainly based on the controlled ITSA results.

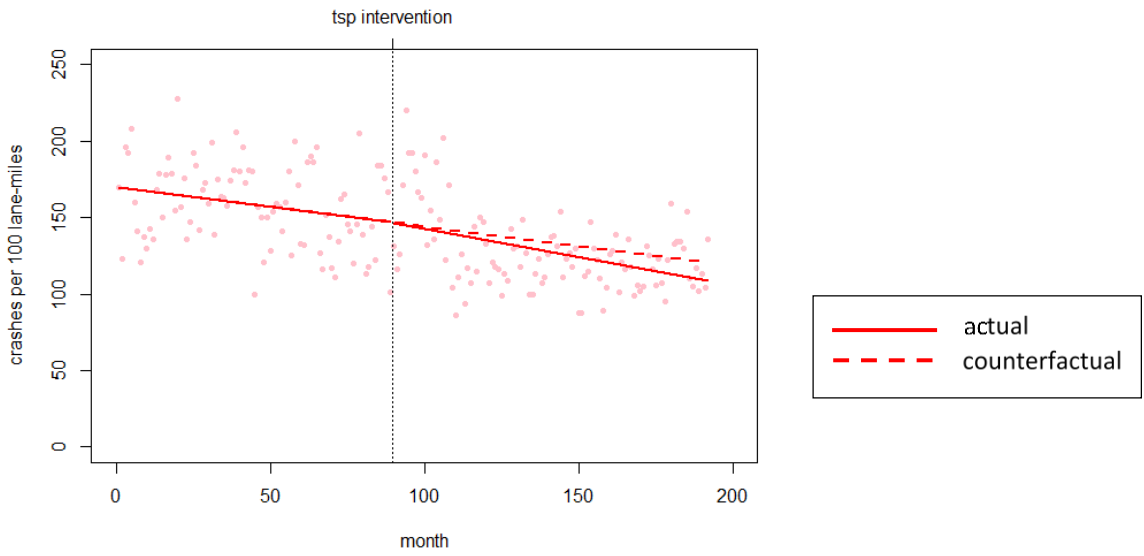

(a) All crashes

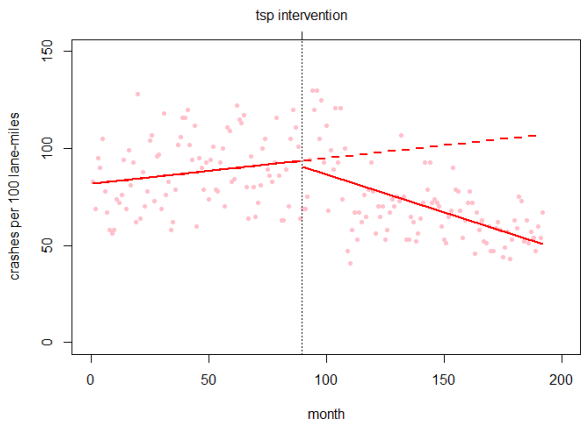

(b) PDO crashes

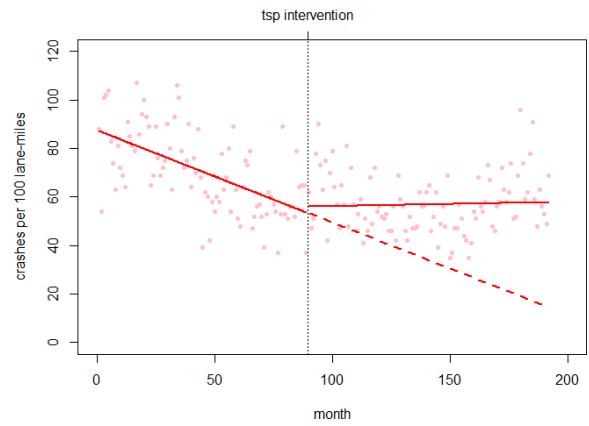

(c) FI crashes

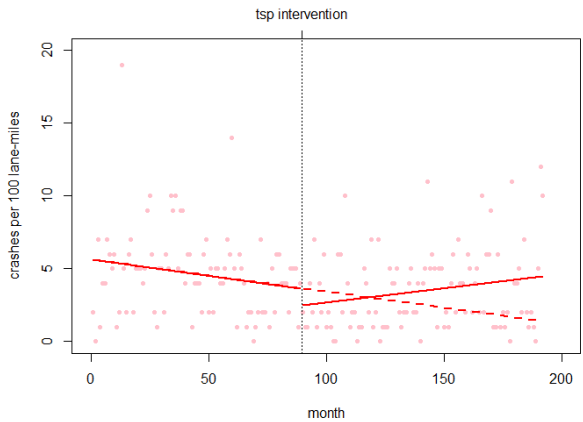

(d) Pedestrian-involved crashes

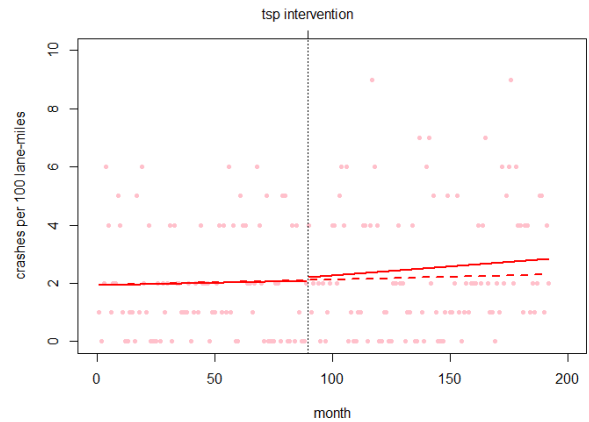

(e) Bike-involved crashes

**Figure 3  Visual Display of Single-Group ITSA Regressions**

**Table 3  Single-Group ITSA Regressions**

| Variable | Coefficient | 95% CI[b] | | p-value | Variable | Coefficient | 95% CI | | p-value | Variable | Coefficient | 95% CI | | p-value |
|---|---|---|---|---|---|---|---|---|---|---|---|---|---|---|
| **All Crashes (AR[1][a])** | | | | | **PDO Crashes (AR[3])** | | | | | **FI Crashes (AR[1])** | | | | |
| *(Intercept)* | 170.92 | 156.01 | 185.84 | 0.00 | *(Intercept)* | 82.65 | 72.19 | 93.10 | 0.00 | *(Intercept)* | 88.35 | 81.19 | 95.50 | 0.00 |
| *time* | -0.25 | -0.54 | 0.03 | 0.08 | *time* | 0.13 | -0.07 | 0.33 | 0.20 | *time* | -0.38 | -0.52 | -0.25 | 0.00 |
| *level* | -0.44 | -21.92 | 21.04 | 0.97 | *level* | -2.77 | -17.85 | 12.32 | 0.72 | *level* | 2.42 | -7.86 | 12.70 | 0.65 |
| *trend* | -0.11 | -0.48 | 0.26 | 0.55 | *trend* | -0.52 | -0.78 | -0.26 | 0.00 | *trend* | 0.40 | 0.22 | 0.57 | 0.00 |
| $M_9$ | -15.43 | -26.41 | -4.45 | 0.01 | $M_9$ | -9.76 | -16.79 | -2.73 | 0.01 | $M_2$ | -7.20 | -13.11 | -1.29 | 0.02 |
| **AR parameter** | | | | | **AR parameters** | | | | | **AR parameter** | | | | |
| *$\phi = 0.36$* | | | | | *$\phi_1 = 0.33; \phi_2 = 0.23; \phi_3 = -0.17$* | | | | | *$\phi = 0.29$* | | | | |

| Variable | Coefficient | 95% CI | | p-value | Variable | Coefficient | 95% CI | | p-value |
|---|---|---|---|---|---|---|---|---|---|
| **Pedestrian-Involved Crashes** | | | | | **Bike-Involved Crashes** | | | | |
| *(Intercept)* | 6.19 | 5.02 | 7.36 | 0.00 | *(Intercept)* | 1.19 | 0.37 | 2.00 | 0.00 |
| *time* | -0.02 | -0.04 | 0.00 | 0.05 | *time* | 0.00 | -0.01 | 0.02 | 0.80 |
| *level* | -1.20 | -2.84 | 0.43 | 0.15 | *level* | 0.10 | -1.03 | 1.23 | 0.86 |
| *trend* | 0.04 | 0.01 | 0.07 | 0.00 | *trend* | 0.00 | -0.02 | 0.02 | 0.66 |
| $M_4$ | -1.56 | -2.98 | -0.13 | 0.03 | $M_5$ | 1.21 | 0.22 | 2.20 | 0.02 |
| $M_6$ | -2.29 | -3.71 | -0.86 | 0.00 | $M_7$ | 0.95 | -0.04 | 1.94 | 0.06 |
| $M_7$ | -1.59 | -3.01 | -0.16 | 0.03 | $M_8$ | 2.63 | 1.64 | 3.62 | 0.00 |
| $M_8$ | -1.39 | -2.82 | 0.03 | 0.06 | $M_9$ | 2.75 | 1.76 | 3.74 | 0.00 |
| | | | | | $M_{10}$ | 1.31 | 0.32 | 2.30 | 0.01 |

Note: a. first-order autoregression; b. 95% confidence interval

**Controlled Analysis**

The overall annual crash numbers of the control group followed a reducing trend over the period from 1995 to 2010. The number of all crashes reduced by 7% (1148 to 1070), where the PDO crashes reduced by 9% (590 to 538), and FI crashes reduced by 5% (558 to 532). The pedestrian-involved crashes increased by 79% (14 to 25). The bike-involved crashes increased by 475% (4 to 23).

Visual displays of the controlled ITSA regressions are presented in Figure 4, with detailed model coefficients and parameters shown in Table 4. In the controlled models, the coefficient estimations for "tsp_level" and "tsp_trend" reflect the immediate (difference in crash number) and temporal (difference in slope) effects on the treated group's monthly crash numbers from the TSP intervention, comparing with the control group, while the coefficient estimations for "level" and "trend" describe the immediate and temporal changes of the control group's monthly crash numbers after the intervention time point. Counterfactual trend for the treated group was assumed to be the same with the control group. Comparing with the counterfactual trend, after the TSP implementations, there was an increase in level of all crashes (9.55, 95% CI [-9.67, 28.77]) with a tiny increase in trend (0.02, 95% CI [-0.46, 0.49]). The PDO crashes had an increase in level (11.25, 95% CI [-9.65, 32.14]) but a decreasing trend (-0.20 95% CI [-0.56, 0.15]). The FI crashes slightly dropped in level (-1.39, 95% CI [-4.95, 16.48]) but had an increasing trend (0.23, 95% CI [-0.12, 0.57]). The pedestrian-involved crashes dropped in level (-0.46, 95% CI [-4.93, 2.01]) and an increasing trend (0.06, 95% CI [0.00, 0.12]). The bike-involved crashes went up in both level (0.38, 95% CI [-1.53, 2.30]) and trend (0.03, 95% CI [0.00, 0.06]). There are significant seasonality variables in all five models. All seasonality variables had a negative sign except for the ones in FI and bike-involved crash models.

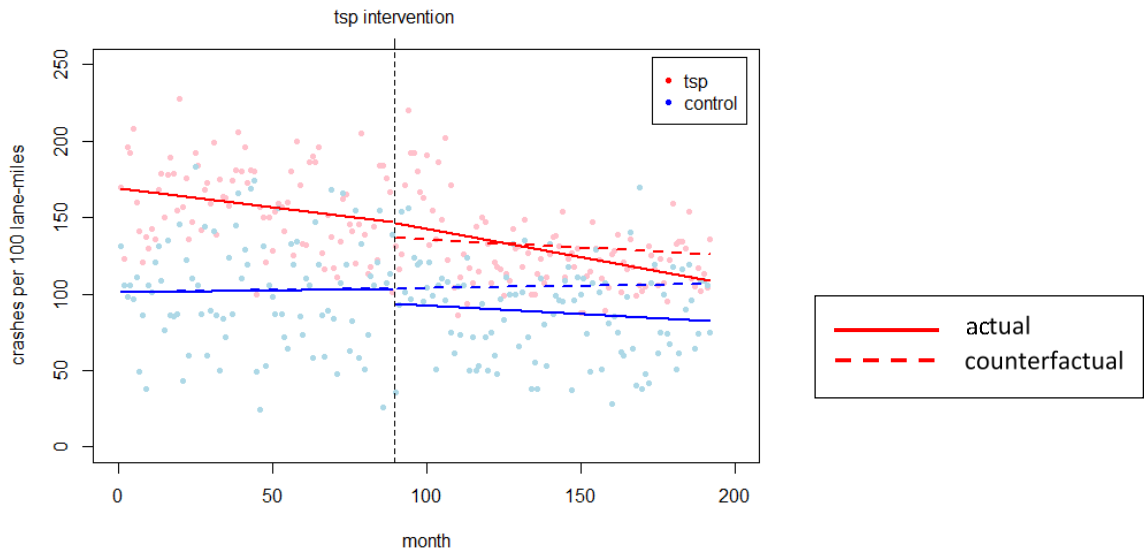

(a) All crashes

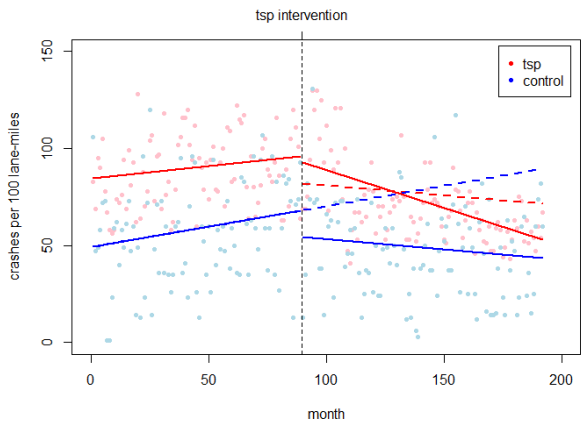

(b) PDO crashes

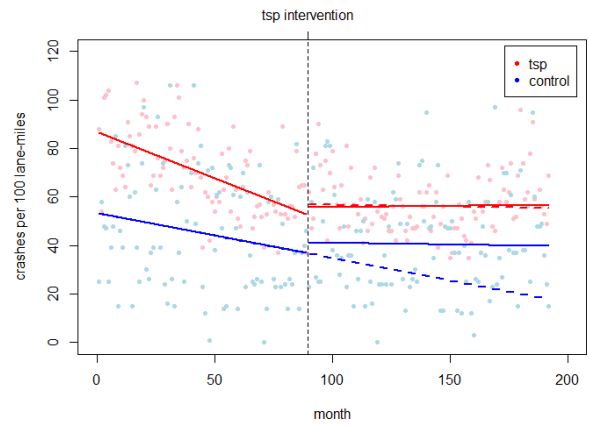

(c) FI crashes

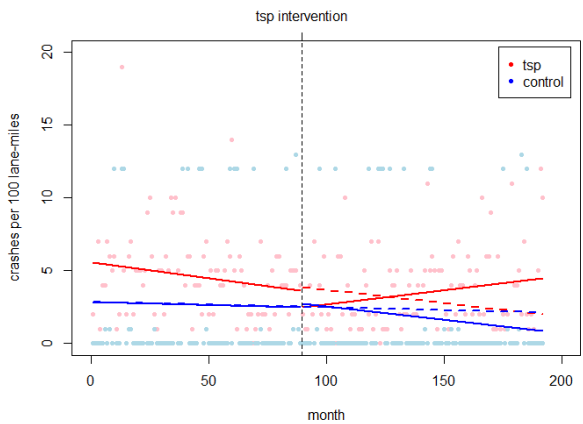

(d) Pedestrian-involved crashes

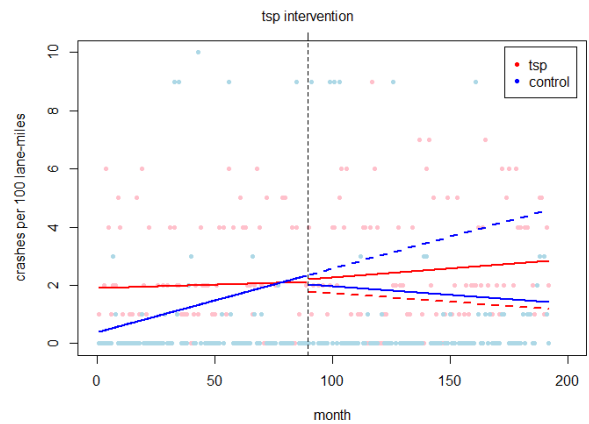

(e) Bike-involved crashes

**Figure 4  Visual Display of Controlled ITSA Regressions**

**Table 4  Controlled ITSA Regressions**

| All Crashes (AR[1][a]) | | | | PDO Crashes (AR[1]) | | | | FI Crashes (AR[5]) | | | |
|---|---|---|---|---|---|---|---|---|---|---|---|
| **Variable** | **Coefficient** | **95% CI**[b] | | **p-value** | | | | | | | |

| **Variable** | **Coefficient** | **95% CI**[b] | | **p-value** | **Variable** | **Coefficient** | **95% CI** | | **p-value** | **Variable** | **Coefficient** | **95% CI** | | **p-value** |
|---|---|---|---|---|---|---|---|---|---|---|---|---|---|---|
| *(Intercept)* | 102.00 | 88.40 | 115.60 | 0.00 | *(Intercept)* | 50.00 | 39.64 | 60.36 | 0.00 | *(Intercept)* | 52.75 | 42.89 | 62.61 | 0.00 |
| *time* | 0.03 | -0.23 | 0.28 | 0.83 | *time* | 0.21 | 0.02 | 0.41 | 0.03 | *time* | -0.18 | -0.37 | 0.00 | 0.05 |
| *tsp* | 68.60 | 49.40 | 87.81 | 0.00 | *tsp* | 35.21 | 20.69 | 49.72 | 0.00 | *tsp* | 33.71 | 19.81 | 47.62 | 0.00 |
| *tsp_time* | -0.28 | -0.64 | 0.09 | 0.13 | *tsp_time* | -0.08 | -0.36 | 0.19 | 0.55 | *tsp_time* | -0.20 | -0.46 | 0.06 | 0.14 |
| *level* | -10.04 | -29.58 | 9.50 | 0.31 | *level* | -13.75 | -28.52 | 1.02 | 0.07 | *level* | 4.30 | -9.80 | 18.40 | 0.55 |
| *trend* | -0.14 | -0.47 | 0.20 | 0.42 | *trend* | -0.31 | -0.57 | -0.06 | 0.02 | *trend* | 0.17 | -0.07 | 0.41 | 0.17 |
| *tsp_level* | 9.31 | -18.32 | 36.94 | 0.51 | *tsp_level* | 11.25 | -9.65 | 32.14 | 0.29 | *tsp_level* | -1.39 | -21.33 | 18.54 | 0.89 |
| *tsp_trend* | 0.02 | -0.46 | 0.49 | 0.94 | *tsp_trend* | -0.20 | -0.56 | 0.15 | 0.27 | *tsp_trend* | 0.23 | -0.12 | 0.57 | 0.20 |
| $M_9$ | -12.72 | -23.06 | -2.38 | 0.01 | $M_3$ | -7.82 | -15.10 | -0.53 | 0.04 | $M_5$ | 7.65 | 1.13 | 14.17 | 0.02 |
| | | | | | $M_6$ | -6.77 | -14.25 | 0.72 | 0.08 | $M_8$ | 8.47 | 1.96 | 14.99 | 0.01 |
| | | | | | $M_7$ | -13.88 | -21.37 | -6.39 | 0.00 | | | | | |
| | | | | | $M_9$ | -8.86 | -16.15 | -1.58 | 0.02 | | | | | |

**AR parameter**

$\phi = 0.11$

**AR parameter**

$\phi = 0.17$

**AR parameters**

$\phi_1 = 0.10$; $\phi_2 = 0.00$; $\phi_3 = 0.05$; $\phi_4 = -0.03$; $\phi_5 = 0.11$

| Pedestrian-Involved Crashes | | | | Bike-Involved Crashes (AR[8]) | | | | |
|---|---|---|---|---|---|---|---|---|

| **Variable** | **Coefficient** | **95% CI** | | **p-value** | **Variable** | **Coefficient** | **95% CI** | | **p-value** |
|---|---|---|---|---|---|---|---|---|---|
| *(Intercept)* | 3.41 | 1.68 | 5.14 | 0.00 | *(Intercept)* | -0.18 | -1.14 | 0.78 | 0.72 |
| *time* | 0.00 | -0.04 | 0.03 | 0.84 | *time* | 0.02 | 0.00 | 0.04 | 0.02 |
| *tsp* | 2.77 | 0.36 | 5.18 | 0.03 | *tsp* | 1.52 | 0.18 | 2.86 | 0.03 |
| *tsp_time* | -0.02 | -0.06 | 0.03 | 0.44 | *tsp_time* | -0.02 | -0.04 | 0.01 | 0.13 |
| *level* | 0.25 | -2.21 | 2.70 | 0.84 | *level* | -0.30 | -1.65 | 1.06 | 0.67 |
| *trend* | -0.01 | -0.06 | 0.03 | 0.50 | *trend* | -0.03 | -0.05 | 0.00 | 0.02 |
| *tsp_level* | -1.46 | -4.93 | 2.01 | 0.41 | *tsp_level* | 0.38 | -1.53 | 2.30 | 0.70 |
| *tsp_trend* | 0.06 | 0.00 | 0.12 | 0.07 | *tsp_trend* | 0.03 | 0.00 | 0.06 | 0.06 |
| $M_6$ | -2.50 | -4.01 | -0.99 | 0.00 | $M_7$ | 1.51 | 0.47 | 2.55 | 0.00 |
| $M_7$ | -1.46 | -2.97 | 0.06 | 0.06 | $M_8$ | 1.85 | 0.82 | 2.88 | 0.00 |
| $M_8$ | -1.70 | -3.21 | -0.19 | 0.03 | $M_9$ | 3.35 | 2.32 | 4.39 | 0.00 |
| $M_9$ | -1.44 | -2.95 | 0.07 | 0.06 | | | | | |

**AR parameters**

$\phi_1 = -0.06$; $\phi_2 = 0.02$; $\phi_3 = -0.12$; $\phi_4 = -0.10$; $\phi_5 = -0.06$;
$\phi_6 = 0.01$; $\phi_7 = -0.04$; $\phi_8 = 0.11$

Note: a. first-order autoregression; b. 95% confidence interval

The effects of TSP can be evaluated over any period after the intervention by calculating these two measures:

$\delta = \lambda - \pi$, the change in expected number of crashes in the post-intervention period; and

$\gamma = \delta / \pi$, the percentage of change in expected number of crashes in the post-intervention period.

Where:

$\lambda$ = the expected number of crashes of the entity in the post-intervention period. This value was estimated; and

$\pi$ = the expected number of crashes of a specific entity in the post-intervention period had the treatment not been implemented. This value was predicted as a counterfactual.

**Table 5  Evaluation of TSP's Effects**

| Year | All | | PDO | | FI | | Pedestrian | | Bike | |
|------|------|------|------|------|------|------|------|------|------|------|
| | $\delta$ | $\gamma$ | $\delta$ | $\gamma$ | $\delta$ | $\gamma$ | $\delta$ | $\gamma$ | $\delta$ | $\gamma$ |
| *2003* | 60 | 3.7% | 52 | 5.3% | 6 | 0.9% | -18 | -35.9% | 14 | 97.6% |
| *2004* | 23 | 1.4% | 11 | 1.1% | 10 | 1.5% | -13 | -26.6% | 15 | 117.7% |
| *2005* | -15 | -0.9% | -31 | -3.2% | 14 | 2.1% | -7 | -16.2% | 17 | 140.6% |
| *2006* | -52 | -3.3% | -72 | -7.7% | 18 | 2.7% | -2 | -4.6% | 19 | 166.9% |
| *2007* | -89 | -5.7% | -113 | -12.4% | 22 | 3.3% | 3 | 8.6% | 21 | 197.6% |
| *2008* | -127 | -8.1% | -155 | -17.2% | 26 | 4.0% | 9 | 23.5% | 22 | 233.8% |
| *2009* | -164 | -10.6% | -196 | -22.2% | 30 | 4.6% | 14 | 40.5% | 24 | 277.1% |
| *2010* | -202 | -13.2% | -238 | -27.3% | 34 | 5.2% | 20 | 60.3% | 26 | 329.8% |
| *Overall* | -565 | -4.5% | -742 | -10.0% | 162 | 3.0% | 6 | 1.8% | 158 | 181.7% |
| | | | | | | | | | | |
| $CRF^a$ | 0.05 | | 0.10 | | -0.03 | | -0.06 | | -1.95 | |
| $CMF^b$ | 0.95 | | 0.90 | | 1.03 | | 1.06 | | 2.95 | |

Note: a. crash reduction factor = -(average of $\gamma$); b. crash modification factor = 1-CRF

The evaluation of TSP's post-intervention effects on crash numbers are listed in Table 5. Crashes were predicted and estimated by month and aggregated by year, to get an understanding of how TSP affected traffic safety over time. By the end of 2002, the TSP systems on the selected 13 treated street sections were all installed and activated, thus the evaluation started from 2003 to the end of the study period, 2010. Overall, the number of all crashes reduced by 4.5%, where PDO crashes reduced by 10.0% and FI crashes increased by 3.0%. Pedestrian-involved crashes increased by 1.8% and bike-involved crashes increased by 181.7%. The overall trend of traffic crashes was decreasing after TSP implementations, thanks to the significant reduction in PDO crashes over time. The FI crashes, pedestrian, and bike-involved crashes, on the other hand, had increasing trends. The negative average of $\gamma$ is equivalent to the crash reduction factor (CRF, with positive value meaning reduction and negative value meaning increase), therefore, (1-CRF) is equivalent to the crash modification factor (CMF). The estimated CMFs of TSP are 0.95 for all crashes, 0.90 for PDO crashes, 1.03 for FI crashes, 1.06 for pedestrian-involved crashes, and 2.95 for bike-involved crashes.

**Discussion**

In this study, safety effects of TSP were assessed using ITSA method. Historical crash data, roadway information, and transit data of Portland, Oregon from 1995 to 2010 were obtained and used in this study. Thirteen bus corridors having TSP installed and activated in June to December 2002 were evaluated using single-group and controlled ITSA. All crashes, PDO crashes, FI crashes, pedestrian-involved crashes, and bike-involved crashes were assessed.

The overall effects of TSP were an initial increase in level and an over-time decrease in trend of crash numbers. Possible reasons to such an outcome could be that the initial confusion or unfamiliarity of road users toward TSP's adjustments to signal timing may lead to a more heterogeneous driver behavior and thus disturbances to traffic movements, which then followed by an adaptation process leading to a more homogeneous driver behavior and eventually smoothed traffic flows along transit corridors. As the transit corridors selected for TSP implementation usually have higher traffic volumes than those street sections not equipped with TSP (which are the case for most of the cross streets of TSP corridors), smoother traffic flows (with fewer stop-and-go's which are correlated with rear-end crashes) helped keep crashes away from the major group of road users.

Looking into different severity levels, the PDO crashes followed exactly this pattern, with an initial increase in level but a significant decreasing trend in the post-intervention period. The FI crashes did not change much in level or trend in the post-intervention period according to the regression model specifications. The visual display showed that for both the treated group and control group, the FI crashes, following a steady reduction in the pre-intervention period, had reached the valley bottom and started to level out in the post-intervention period. This change starting at the same time with the TSP intervention was significant in both the treated group and the control group. Therefore, the difference between the leveled-out treated group FI crashes and the leveled-out control group FI crashes, which made up the overall 3.0% increase in the evaluation, was mostly due to the differences in some of the two groups' organic features rather than the direct impact from TSP. These differences were not statistically significant; therefore, these uncertainties were considered acceptable.

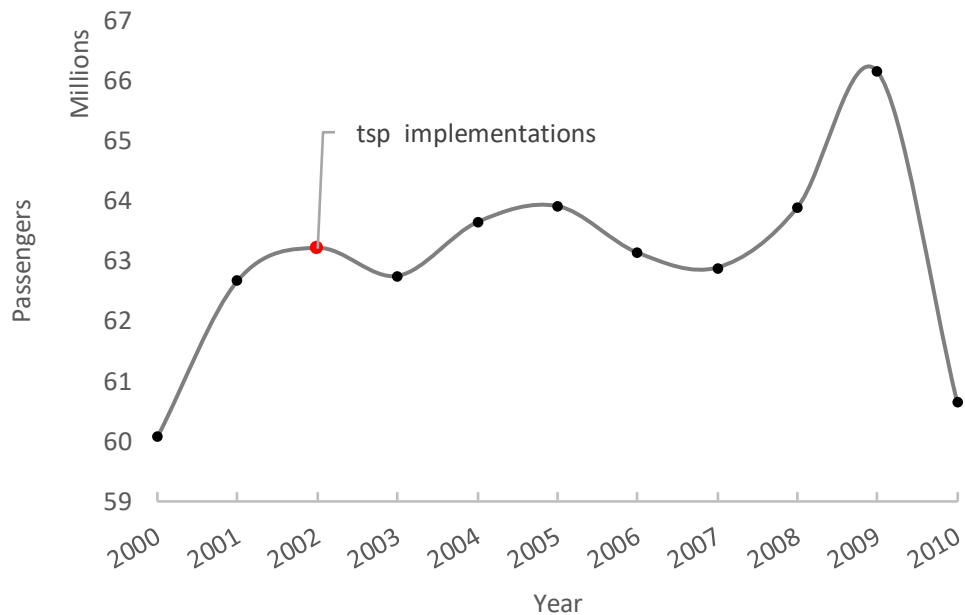

**Figure 5  Portland TriMet Bus Boarding Ridership (2000 to 2010)**

In terms of pedestrian and bike-involved crashes, the ITSA model coefficients and evaluation results indicated an increase after the TSP intervention. However, the small numbers of pedestrian and bike-involved crashes (single digit annual crashes per mile per lane) led to the large percentage of increase in crash numbers. Regarding potential reasons to these increasing trends, note that TSP, as a transit preferential treatment, reduces bus delay thus improves service quality and attractiveness significantly, as indicated in existing studies (*1-4*). Therefore, bus ridership may increase due to the improvement in service attractiveness, which was the case for Portland, as supported by the ridership data (obtained from TriMet website) plotted in Figure 5. It shows that from 2002 to 2010, the TriMet bus boarding ridership followed an overall increasing trend. TSP systems generally ensure minimum pedestrian phases, but bicycles traveling on motor vehicle lanes follow motor vehicle signals, which are adjusted when TSP kicks in. Therefore, the increase in pedestrian-involved crashes may be affected by the increased ridership, but the increase in bike-involved crashes may relate more to the TSP signal adjustments, and other bike-related factors (e.g., increase in bike traffic, as Portland has been progressively promoting biking). Due to the lack of pedestrian and bike traffic volume data as exposure, the quantities of effects from TSP, induced ridership, and increased bike traffic, could not be distinguished by the ITSA models developed in this study.

## CONCLUSIONS

From the above-summarized findings, conclusions can be drawn that the TSP implementations in Portland, Oregon was effective in reducing traffic crashes along transit corridors over time, especially in reducing PDO crashes over time. The FI crashes started leveling out and were not affected by the TSP intervention. Pedestrian and bike-involved crashes increased after TSP intervention but could be caused by increased pedestrian and bicycle traffic, the effects of which could not be distinguished with the current ITSA.

The evaluated effectiveness of TSP in reducing traffic crashes was comparable with the results reported by most of the existing studies. Three Australian studies and a previous study by the authors of this paper, using King County, Washington, data all reported reduction in crashes with TSP implementations (16-19).

The ITSA provides details about the changes in multiple categories of crashes throughout the entire study period, with model estimations and visual displays. The effects of a treatment should be evaluated over time based on a continuous trend, rather than based on one effectiveness estimation over one period. The results from ITSA should be very useful to transportation and transit agencies in evaluating past TSP implementations, as well as planning for future TSP implementations. From the findings of this study, there are two points worth noting by the agencies thinking about implementing TSP. First is that TSP has an effect of reducing the number of crashes, especially PDO crashes, over time. Second is that TSP has a potential effect of increasing pedestrian and bike-involved crashes. However, safety effects of TSP can vary in different places under different conditions and settings, and assessments like this one will be needed before any implementations of such a system.

There are some limitations of this study. First, crash number was used as a surrogate measure for traffic safety. Although crashes seem to be the results which can be linked to infrastructure, traffic, and human factors, this surrogate measure lacks a good knowledge of the preceding dynamics of the incidents, which may provide better understanding of the causes and support the determination of appropriate treatments (44). Moreover, crash data are usually gathered from police reports or hospital records, which sometimes have issues in consistency and availability. The random nature and uniqueness of crashes also bring methodological challenges when modeling them. Second, the ITSA models were fitted with the assumption of linearity. In this study, the pre-intervention linearity of trends over time was checked to be satisfied with visualization and statistical tests, for both the treated and control groups. Seasonality and autoregression was also addressed by adding seasonality variables and autoregressive error terms. More sophisticated ITSA modeling techniques for discrete response variables can be applied if trend check shows significant violation of linearity assumptions. Third, no detailed (vehicular, pedestrian, and bike) traffic volume data were available as exposure measures. Although (number of lanes * street section length) was used as a surrogate exposure, traffic volume would be a better exposure for crash regression modeling. Traffic volume changes over time and has been proven to be significantly correlated with crash frequency.

## ACKNOWLEDGMENTS

The authors would like to thank the Portland Bureau of Transportation, for providing the TSP, traffic, and street data, and TriMet, for providing the transit data, used this research. The authors would also like to thank the Oregon Department of Transportation and Portland State University for maintaining the Oregon Traffic Safety Data Archive. The work presented in this paper remains the sole responsibility of the authors.